\documentclass[twocolumn,showpacs,preprintnumbers,amsmath,amssymb]{revtex4}
\usepackage{epsfig}
\usepackage{graphicx}
\usepackage{dcolumn}
\usepackage{bm}

\def\be{\begin{equation}}
\def\ee{\end{equation}}
\def\bea{\begin{eqnarray}}
\def\eea{\end{eqnarray}}

\def\p{\partial}

\begin{document}

\preprint{}

\title{How to Falsify the GR$+\Lambda$CDM Model \\
with Galaxy Redshift Surveys}

\author{Viviana Acquaviva}
\email{vacquaviva@physics.rutgers.edu}
\author{Eric Gawiser}
\affiliation{Department of Physics and Astronomy, Rutgers, The State University of New Jersey, Piscataway, 08854}

\date{\today}

\begin{abstract}
  A wide range of models describing modifications to General Relativity have been proposed, but no fundamental parameter set exists to describe them.   Similarly, no fundamental theory exists for dark energy to parameterize its potential deviation from a cosmological constant.  This motivates a model-independent search for deviations from the concordance GR$+\Lambda$CDM cosmological model in large galaxy redshift surveys.  We describe two model-independent tests of the growth of cosmological structure, in the form of quantities that must equal one if GR$+\Lambda$CDM is correct.  The first, $\epsilon$, was introduced previously as a scale-independent consistency check between the expansion history and structure growth.  The second, $\upsilon$, is introduced here as a test of scale-dependence in the linear evolution of matter density perturbations.  We show that the ongoing and near-future galaxy redshift surveys WiggleZ, BOSS, and HETDEX will constrain these quantities at the $5-10\%$ level, representing a stringent test of concordance cosmology at different redshifts. When redshift space distortions are used to probe the growth of cosmological structure, galaxies at higher redshift with lower bias are found to be most powerful in detecting the presence of deviations from the GR$+\Lambda$CDM model. However, because many dark energy or modified gravity models predict consistency with GR$+\Lambda$CDM at high redshift, it is desirable to apply this approach to surveys covering a wide range of redshifts and spatial scales.
 \end{abstract}

\pacs{}
\maketitle

\section{Introduction}

Evidence for accelerated cosmic expansion \cite{Riess:1998cb,Perlmutter:1998np}
 poses a serious challenge to the standard paradigm of Big Bang cosmology, where gravity is described by General Relativity (GR), and the Universe is filled by matter and radiation fields. Most proposed explanations invoke failure of GR on cosmological scales or introduce a new component with strongly negative equation of state in the cosmic energy budget. Without compelling theoretical reasons to support any particular theory, Occam's razor favors the model able to explain all of the cosmological observations with the minimum number of parameters. Currently, that model is ``GR+$\Lambda$CDM", characterized by two assumptions: gravity is correctly described by GR, and cosmic acceleration is caused by a cosmological constant ($\Lambda$). We refer to it as the {\it concordance} cosmological model. 
 
The rich literature in this field offers many explanations for the accelerated expansion alternative to the concordance model  (for a list of references, see \cite{Jain:2010ka}). According to which of the two key assumptions 
is broken, they fall in the category of Modifications of Gravity (MoG) or Dark Energy (DE). 
Instead, we consider the concordance model in its entirety as the reference framework and ask ourselves how we can most thoroughly test that all cosmological observations are in agreement with it. 

The motivation for this approach is manifold. First of all, 
it is {\it feasible}. While interpretation of the data in the context of a given theory is necessarily model-dependent, the detection of a significant deviation from an expected behavior is incontrovertible.
 Moreover, the unlimited 
freedom intrinsic in more complicated MoG and DE models is reflected by degeneracies in cosmological observables, so that it is impossible to uniquely identify the phenomenology of one or the other class, without making further theoretical assumptions ({\it e.g.}, \cite{Simpson:2009zj}). 
Second, our approach 
is {\it relevant}. If the concordance model were proven to be faulty, this would profoundly change our understanding of the Universe, whatever the exact explanation might be. 
Finally, this approach is {\it sensible}; practicality suggests that if 
the concordance model is 
falsified it would motivate a far more detailed exploration of 
the possible underlying reasons, a search 
then guided 
by knowledge of the nature, and location in space and time, of the observed deviation. 

The paper is organized as follows. The first step, discussed in Sec. \ref{secI}, is to identify relations between cosmological observables, which hold for the concordance model. This is similar in spirit to what was proposed in \cite{AHSD}, but here we further break down the concordance model behavior in order to obtain two independent relationships. This allows us to identify two parameters, named $\epsilon$ and $\upsilon$, whose value can be exactly predicted for the GR+$\Lambda$CDM case. 
In Sec. \ref{secII} we write these parameters in terms of observable quantities, and we select the methods for measuring the quantities entering the definition of $\epsilon$ and $\upsilon$. We focus on the use of the power spectrum and bispectrum of galaxy redshift surveys, in conjunction with CMB measurements. 
The established framework is used in Sec. \ref{secIII} to define a {\it Figure of Merit} in the $(\epsilon, \upsilon)$ plane for the task at hand. The computation of the Figure of Merit requires knowledge of the observational uncertainties in the measurement of  $\epsilon$, $\upsilon$, and the other relevant cosmological parameters; these are determined using the Fisher Matrix technique. We compute the actual Figure of Merit for the surveys WiggleZ, BOSS and HETDEX. This approach allows a direct model-independent comparison of
the capabilities of different experiments in answering the question ``Is GR$+\Lambda$CDM the correct cosmological model?" We summarize our results and list concluding remarks in Sec. \ref{secIV}.

\section{Theoretical framework}
\label{secI}

We assume scalar linear perturbations around a flat FRW background
in the Newtonian gauge,
\be
ds^2 = -(1+2\Psi) dt^2 + a^2(1-2\Phi)dx^2
\ee
 and work in the quasi-static, linear approximation, which is valid for sub-horizon modes still in the linear regime {\it e.g.}, \cite{Dent:2009wi}. 
 
 The evolution of perturbations is described by the continuity, Euler and Poisson equations ({\it e.g.}, \cite{Jain:2007yk}):
\bea
\delta'_m &=& -\frac{k}{aH} v_m, \\
v'_m + v_m &=& \frac{k}{aH} \Psi, \\
k^2 \Phi  &=& - 4\pi G a^2 \rho_m \delta_m.
\eea
where $\delta_m$ and $v$ are the matter density perturbations and the divergence of the peculiar velocity, respectively. In the concordance model, the anisotropic stress vanishes, and these equations can be combined to derive the equation of motion
for the matter density perturbations, or equivalently for the growth factor $D$ defined as $D(k,a) = \delta_m(k,a)/\delta_m(k,a=0)$:
 \be
\label{growth}
D'' + (2 + \frac{H'}{H})D' - \frac{4\pi G}{H^2}\rho_m D = 0;
\ee
where $a$ is the scale factor, $H = \dot{a}/a$ is the Hubble function, a prime denotes derivative with respect to $\ln a$, and $\rho_m$ is the average matter density.  

Two well-known features of the GR$+\Lambda$CDM model are immediately evident from the form of this equation
({\it e.g.} \cite{AHSD}).
The solution for the linear growth factor can be determined exactly once the Hubble function is known; and, since none of the coefficients is a function of scale, the growth factor is scale-independent. These are two independent properties of the concordance model, and we propose to test them separately
 using the two parameters
\bea
\epsilon(a) & = & \Omega_m^{-\gamma(a)} f(a); \\ 
\label{eq:eps}
\upsilon(a) & = & \frac{f(k_s, a)}{f(k_L,a)}.
\label{eq:ups}
\eea

In the above equations, $\Omega_m$ is the matter contribution to the total density, $f = \frac{d \ln D}{d \ln a}$, $k_{\rm{s(mall)}}$ and $k_{\rm {L(arge)}}$ represent two different scales where the logarithmic derivative of the growth is measured, and $\gamma(a)$ is the simple function entering the well known fitting function $(d \ln D / d \ln a)_{\rm fit}  \simeq [\Omega_m(a)]^{\gamma(a)}$.  Throughout this paper, we use for the fit the 
value $\gamma = 0.55$ \cite{Linder:2005in}, which is accurate at the $0.05\%$ level for GR+$\Lambda$CDM at the redshifts of interest.  

The definition of the parameter $\epsilon$ is similar to the one used in \cite{AHSD} 
but with one important difference: we assume here that the measurement of $f$ in $\epsilon$ is {\it averaged} over all the scales probed by each galaxy redshift survey. Therefore, the parameter $\epsilon$ is only sensitive to the consistency between the Hubble expansion (equivalent to $\Omega_m(a)$) and the growth as a function of time $D(a)$, required by the concordance model, while $\upsilon$ is sensitive to a scale dependence in the logarithmic derivative of the growth. 
If many measurements of the growth as a function of scale are available, $\upsilon$ could be defined in multiple ways. 
We choose to maximize the signal to noise for the measurement of this parameter by using two bins in $k$ for each survey, and choosing the ranges 
so that the two measurements have similar fractional uncertainties.

Both parameters are exactly $= 1$ at all times in the GR$+\Lambda$CDM model, and this formalism allows for an easy detection of deviations from the concordance model behavior, although it would not tell us immediately whether non-$\Lambda$ Dark Energy or 
MoG is the culprit.  

A scale dependence in the 
logarithmic derivative of the growth factor 
 (corresponding to 
 $\upsilon \ne 1$) 
 is challenging within the context of perfect fluid Dark Energy models \cite{Unnikrishnan:2008qe}, but is predicted by models 
 with varying sound speed, or anisotropic stress {\it e.g.,} \cite{DeDeo:2003te,Koivisto:2005mm}. Limits on such quantities are still weak \cite{dePutter:2010vy}. 
Massive neutrinos may also induce a scale-dependent growth factor. Neutrinos are assumed massless in the concordance model. In the case of neutrinos of non-negligible but known mass, the effect of such scale dependence can be removed and the formalism applied. If the same galaxy survey is used to measure the logarithmic growth and the mass of neutrinos, we expect the Figure of Merit to be smaller.
 
\section{From theory to observables}
 \label{secII}
In order to compute a Figure of Merit for the measurement of 
$\epsilon$, and $\upsilon$, we start by rewriting them in terms of observable quantities. We have:
\bea
\epsilon(a) & = & \Omega_m^{-\gamma(a)} f(a) = \frac{a^{3\gamma} H(a)^{2 \gamma} }{(\Omega_{m,0} H_0^2)^{\gamma}} \beta(a) \, b(a);\\
\label{eq:epsobs}
\upsilon(a) & = & \frac{f(k_s, a)}{f(k_L,a)} = \frac{\beta(k_s,a) \, b(a)}{\beta(k_L,a)\,b(a)} = \frac{\beta(k_s,a)}{\beta(k_L,a)} .
\label{eq:upsobs}
\eea
Here, we have written $f$ as $\beta \cdot b$, where $\beta$ is the linear redshift distortion parameter, and $b$ is the bias of the target galaxies. 
We assume that any scale dependence of the bias has been removed; see Sec. \ref{secIII} for further discussion.

CMB experiments are an excellent probe of the combination $\Omega_{m,0} H^2_0$, 
whose value is 
known to within 5\% from WMAP7 data \cite{Larson:2010gs}. 
We assume that this measurement will be performed by the ongoing CMB satellite mission Planck \cite{Planck} with a relative precision of $1.5\%$.
We represent this constraint as a Gaussian prior in the likelihood function.  

The linear bias of galaxies can be measured in 
several ways; here we assume it has been measured via the bispectrum of galaxies, as suggested in \cite{Sefusatti:2007ih}. That paper reports the precision
for the bias measurement achievable by several reference surveys, together with a discussion on how the precision varies as a function of survey parameters ({\it e.g.}, number density of galaxies, volume, $k_{\rm max}$). We use simple scaling relations, derived from their discussion, in order to determine how well galaxy surveys 
can measure the bias, and again introduce this constraint in the likelihood function as a Gaussian prior. We assume that no information on the bias is derived from the power spectrum, since a correct implementation of a simultaneous measurement of the bias from bispectrum and power spectrum would require knowledge of the covariance between them. We note that as a result, we might be underestimating the potential of a given survey in measuring the bias. \\
The remaining quantities appearing in the definition of $\epsilon$ and $\upsilon$ are the Hubble parameter at a given time, $H(a)$, and the linear redshift distortion parameters measured on two different scales, $\beta(k_s,a)$ and $\beta(k_L,a)$. 
These functions can be measured by galaxy  redshift surveys. It was shown in \cite{Shoji:2008xn} that fitting the full 2-D galaxy power spectrum allows one to maximize the signal to noise in the measurement of $H(a)$ and $\beta$. We modify 
the public Fisher Matrix code available at \cite{Eichiirospage}
to account for the scale dependence of the linear redshift distortion parameter and to adopt the new parameterization in terms of $\epsilon$ and $\upsilon$.  \\

\section{Figure of Merit} 
\label{secIII}
We now perform a Fisher Matrix calculation 
and use the results 
to generate a Figure of Merit in the plane ($\epsilon$, $\upsilon$). We assume that each survey provides a measurement of the relevant parameters at a single redshift (typically 
the mean redshift of the survey), and therefore we drop the time dependence of the parameters.
This allows one to directly compare the performance of different surveys, even when they target different redshift ranges.
For notation 
convenience, we 
denote $\beta(k_s)$ and $\beta(k_L)$ as $\beta_s$ and $\beta_L$ respectively. 

As a first step, 
we compute the Fisher Matrix for the set of parameters $ p_i = \{ \beta_s, \beta_L, \ln{H}, \ln{A}, n_s, b, \sigma_v, \Omega_{m,0} H^2_0 \}$; $A$ and $n_s$ represent the amplitude and scalar spectral index of the primordial perturbations, and $\sigma_v$ is the peculiar velocity dispersion we use to model the Finger of God effect \cite{Jackson:2008yv}, following \cite{Ballinger:1996cd,Shoji:2008xn}. We use the measurement of the galaxy power spectrum in redshift space for the first 
six parameters, and assume no covariance between them and the measurements of the bias 
(obtained from bispectrum) 
and $\Omega_{m,0} H_0^2$ 
(obtained from the CMB), as 
described 
above.

For a galaxy redshift survey, the Fisher Matrix has the form ({\it{e.g.}}, \cite{Seo:2003pu})
\be
\label{eq:FM}
F_{ij} = \int^{k_{\rm max}}_{k_{\rm min}} \frac{4 \pi k^2 d k}{(2 \pi)^3} \int^1_0 d\mu \frac{\partial \ln P_g(k, \mu)}{\partial p_i}\frac{\partial \ln P_g(k, \mu)}{\partial p_j} w(k,\mu).
\ee
Here 
$\mu$ is the cosine of the angle between the vector $  k$ and the line of sight, $w(k,\mu)$ is the weight function
\be
w(k, \mu) = \frac{1}{2} \left[ \frac{n_g P_g(k, \mu)}{1 + n_g P_g(k, \mu)} \right] V, 
\ee
 accounting for the survey's volume and number density of galaxies $n_g$,
and $k_{\rm max}$ is the maximum wavenumber for which the formalism is valid \footnote{We assume $k_{\rm max} = 0.1 h$ Mpc$^{-1}$ at $z=0$, and rescale it to the other redshifts by  
holding $\delta_{k_{\rm max}}$ constant.  
}.
We model the linear part of $P_g(k, \mu)$ analytically as in \cite{Shoji:2008xn}:
\be
P_g(k,\mu) = b^2 [P_{\delta \delta}(k) + 2 \beta \mu^2 P_{\delta v}(k) + \beta^2 \mu^4 P_{v v}(k)];
\ee
$\delta$ refers to density and $v$ to velocity perturbations.

Many of the relevant derivatives appearing in the Fisher Matrix formula (\ref{eq:FM}) are given in the appendix of \cite{Shoji:2008xn}, and we do not report them here. The newly introduced derivatives with respect to $\beta_s$ and $\beta_L$ can be obtained using the same formula as the derivative with respect to $\beta$, but integrating in two separate $k$ bins. If $k_{\rm med}$ is the wavenumber chosen to separate the measurement of ``small'' and ``large'' scales, 
\bea
\frac{d \ln P_g(k)}{d \beta_s} & = &  \theta(k, k_{\rm med})\frac{d \ln P(k)}{d \beta}; \\
\frac{d \ln P_g(k)}{d \beta_L} & = & (1- \theta(k, k_{\rm med}))\frac{ d \ln P(k)}{d \beta};
\eea
where $\theta(k_1, k_2) = 0$ if $k_1 < k_2$, and $1$ otherwise. 

The second step is to transform to the final set of parameters $q_i = \{ \epsilon, \upsilon, \ln{H}, \ln{A}, n_s, b, \sigma_v, \Omega_{m,0} H_0^2 \}$, and compute the corresponding Fisher Matrix, $\tilde{F}_{a b}$:
\be
\label{eq:jac}
\tilde{F}_{a b} = \sum_{ij} \frac{\partial p_i}{\partial q_a} \frac{\partial p_j}{\partial q_b} F_{ij}.
\ee
For this purpose, it is useful to rewrite the parameters $\epsilon$ and $\upsilon$ explicitly in terms of $\beta_s$, $\beta_L$, $H$, $b$ and $\Omega_{m,0} H_0^2$:
\bea
\epsilon & = & \frac{a^{3\gamma} H^{2 \gamma} }{(\Omega_{m,0} H_0^2)^{\gamma}} \frac{\beta_s + \beta_L}{2} \,b ;\\
\upsilon & = & \frac{\beta_s}{\beta_L},
\eea
where we have used the fact that the parameter $\beta$ appearing in $\epsilon$ is an average of the measurements over small and large scales, $a$ is the scale factor at the mean redshift of each survey, and $\gamma$ is the corresponding value of the fitting function introduced earlier. For the concordance model, $\beta_s = \beta_L = \beta$.

The non-null derivatives appearing in Eq. (\ref{eq:jac}) are:

\bea
\frac{\p \beta_s}{\p \epsilon} & \quad = \quad & \frac{\p \beta_L}{\p \epsilon} = \frac{2 \left( \frac{a^3 H^2}{\Omega_{m, 0} H_0^2}\right)^{- \gamma}}{b}; \\
\frac{\p \ln H}{\p \epsilon}  &  \quad  =  \quad  & \frac{\left( \frac{a^3 H^2}{\Omega_{m, 0} H_0^2}\right)^{- \gamma}}{b\, \gamma (\beta_s + \beta_L)}; \\
\frac{\p (\Omega_{m, 0} H_0^2)}{\p \epsilon}  &  \quad =  \quad & - \frac{2 \left( \frac{a^3 H^2}{\Omega_{m, 0} H_0^2}\right)^{- \gamma} \Omega_{m, 0} H_0^2}{b\, \gamma (\beta_s + \beta_L)}; \\
\frac{\p b}{\p \epsilon}  &  \quad  =  \quad & \frac{2 \left( \frac{a^3 H^2}{\Omega_{m, 0} H_0^2}\right)^{- \gamma}}{\beta_s + \beta_L}; \\
\frac{\p \beta_s}{\p \upsilon} &  \quad =  \quad & \beta_L ;\\
\frac{\p \beta_L }{\p \upsilon} &  \quad =  \quad & - \frac{\beta^2_L}{\beta_s}.
\eea  
\begin{table}[t]
  \center
  {\small
  \begin{tabular}{lccccccc} 
    \hline
    \hline
 Galaxy & \quad $N_{\rm gal}$ & $\quad { V} $ & $z_m$ & $k_{\rm med}$ &$k_{\rm max} $ & $b$ & $ \; \;\,\Delta b/b$\\
 Survey     &  &  {\tiny $(h^{-3} {\rm Gpc}^3)$} & & \tiny $(h / \rm{Mpc})$ & \tiny $(h / \rm{Mpc})$ & &  \\
    \hline
     \hline

     WiggleZ      & $\,0.24\cdot10^6$   &   1.0 &  0.6 &0.1& 0.2 & 1  &  0.06\\

     \hline

    \hline

     BOSS      & $\,1.5\cdot10^6$   &   4.4 &  0.45 & 0.1 & 0.18 & 2  &  0.03\\

     \hline    

     HETDEX  &  $\,0.8\cdot10^6$    &   3.0  &  2.7  &  0.15 & 0.4 & 2 &  0.015\\
     \hline
     \hline
  \end{tabular}
  }
\caption{\label{tab:par} Specifics of the considered surveys. 
We display the total number of galaxies $N_{\rm gal}$, the volume $V$, the mean redshift $z_m$, the wavenumber $k_{\rm med}$ where we split the surveys, the maximum 
linear
wavenumber $k_{\rm max}$, the galaxy bias $b$, and the percentage error in the bias.}
\end{table}

With these derivatives in hand, we can now compute the Figure of Merit for three representative galaxy surveys, WiggleZ (ongoing, \cite{WiggleZ}), BOSS (ongoing, \cite{BOSS}) and HETDEX (planned, \cite{HETDEX,Hill:2008mv}), targeting respectively blue emission-line galaxies ({\it e.g.}, \cite{Ellis:1997td}), Luminous Red Galaxies ({\it e.g.}, \cite{Eisenstein:2001cq}) and Lyman Alpha Emitters ({\it e.g.}, \cite{Gawiser:2007vm}). The assumed specifics of the surveys are reported in Table \ref{tab:par}; the value of $\Delta b$ is the $1\sigma$ amplitude of the Gaussian prior on the linear bias. For WiggleZ, we assume the final size of the survey as a reference.

\begin{figure}[t]
\includegraphics[width=8.5cm]{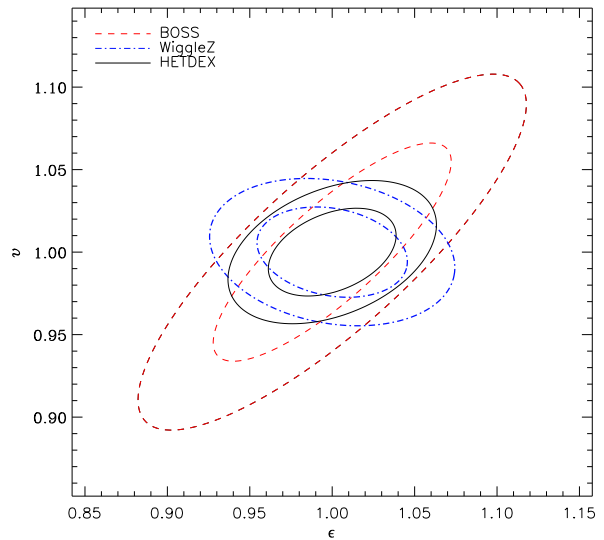}
\caption{Expected errors in the $(\epsilon, \upsilon)$ plane for the 
WiggleZ,
BOSS and 
HETDEX
surveys.  The concentric ellipses show the 1 and 2 $\sigma$ contours. The values of the Figure of Merit, defined as $\pi/A$, where A is the area of the 1-$\sigma$ ellipse,
are 204, 93 and 262, respectively.}

\label{fig:ellip}
\end{figure}

The marginalized errors in the $(\epsilon, \upsilon)$ plane for the three surveys are shown in Figure \ref{fig:ellip}. While the three surveys are exploring different redshift windows, it is fair to plot the constraints on the same plane, because $\epsilon$ and $\upsilon$ are constructed to be exactly equal to one for the concordance model at all redshifts. 
The inverse of the area of the ellipse thus provides a Figure of Merit that 
allows a fair comparison of the strength of different surveys in detecting deviations from the concordance model. However, 
constraints centered at different redshifts are independent from each other. 

Figure \ref{fig:ellip} shows that HETDEX will perform slightly better than WiggleZ in detecting deviations from the GR$+\Lambda$CDM model, and both will improve significantly on the result from BOSS. More specifically, the $1\sigma$ uncertainty 
in the measurement of $\epsilon$, $\upsilon$ is 
6\%,4\%
 for WiggleZ,
10\%,  9\%
for BOSS, 
and 5\%,4\% for HETDEX.
The corresponding values of the Figure of Merit, defined as $\pi/A$, are 204, 93 and 262 respectively.
 
The results mainly depend on the following factors.  First, the measurement of the linear bias obtained via the bispectrum is noticeably better at higher redshift, as evident from Figs. 5 and 6 in \cite{Sefusatti:2007ih}. Higher-redshift surveys are also favored by the larger number of available modes in the linear regime.
Similarly, the signature of $d \ln D / d \ln a$ (equivalent to $\beta$ for galaxies of same bias) is larger at $z \simeq 2 $ than $z \simeq 0.5$, because the growth is more rapid during matter domination than at later times, when the dark energy slows down the growth of the perturbations. These three factors reduce the uncertainties in the measurement of $\epsilon$ and $\upsilon$, in the case of HETDEX vs. BOSS. 
Finally, for a similar rate of growth, the linear redshift distortion signal is largest for galaxies of low bias (such as the ones targeted by WiggleZ). This not only positively influences the measurement of $\beta_s$ and $\beta_L$, but also alters the correlation between $\epsilon$ and $\upsilon$. For the particular configuration of WiggleZ, $\epsilon$ and $\upsilon$ are less correlated than for the other two surveys. This effect is slightly mitigated by the worse sensitivity of WiggleZ to $\sigma_v$ and $\beta$, resulting from the smaller volume and bias factor, but the resulting Figure of Merit is still exceptionally promising. 
One possible caveat about the use of the lowest-bias galaxies comes from the N-body simulations performed by the authors of \cite{Okumura:2010sv}. We have assumed here that the scale dependence of the bias is either absent, or it is possible to correct for it. Their work shows that the bias is reasonably independent of scale in the most massive halos, where high-bias galaxies reside, but not in low-mass halos, where low-redshift Emission Line Galaxies are. If accounting for such scale dependence turns out to be complicated, this would favor surveys like HETDEX and BOSS versus WiggleZ. 

Finally, the choice of a different split in $k$ space mildly affects the resulting Figure of Merit and the correlation between $\epsilon$ and $\upsilon$, as long as the achieved precisions in the two bins are not very different (in which case the uncertainties become larger). One could choose $k_{\rm med}$ in different ways, for example choosing the one that minimizes the FoM or the correlation between the two parameters. However, this optimization works for one survey at a time. The prescription used here to determine $k_{\rm med}$  allows for a simple and fair comparison of multiple surveys.

For the same reason, we chose to focus in this paper on the minimal number of bins in wavenumber and in redshift that allow both parameters to be measured. The combined power of surveys at different redshifts to discriminate against GR+$\Lambda$CDM can be easily obtained by adding their Fisher Matrices. There are obvious advantages in having a broad lever arm in both the radial ($z$) and transverse ($k$) direction; if the concordance model were falsified, breaking down the information might reveal deviations in only one bin, and cast light on the nature of such deviation. However, looking at probes at multiple redshifts and/or several bins in $k$ is survey-specific and is beyond the scope of this paper.

\section{Conclusions}
\label{secIV}
We have shown how redshift galaxy surveys can be used to test the consistency of the concordance GR+$\Lambda$CDM model with the observations \cite{Wang:2007ht,Song:2008xd,Song:2008vm,Percival:2008sh,White:2008jy}. 

This approach is motivated by the idea that the simplest model should be abandoned only if evidence of its failure is found. This framework allows one to search for deviations from the concordance model behavior in a completely model-independent fashion. The two features of the concordance model we explored are the fixed relationship between the expansion history and the logarithmic derivative of the growth, and the fact that the latter is independent of scale. Each of these relationships can be rewritten in terms of a parameter whose value is known in the concordance model. We have defined these parameters as $\epsilon$ and $\upsilon$, and used a Fisher Matrix formalism to derive a Figure of Merit for three galaxy surveys, WiggleZ, BOSS and HETDEX, in the $(\epsilon, \upsilon)$ plane. For this particular choice of observables and techniques ({\it e.g.}, the use the bispectrum to determine the bias), and assuming that any scale dependence of bias can be removed,
high-redshift, low-bias galaxies are found to be more powerful in detecting the presence of deviations from the concordance model. However, if such deviations were found, characterizing them would require a comprehensive mapping of the expansion history and growth of structure on multiple redshifts and scales.

The framework we developed for testing the GR+$\Lambda$CDM model can be easily applied to other measurements of $\epsilon$ and $\upsilon$; many of current constraints on the growth of structure 
\cite{Hawkins:2002sg,Verde:2001sf, Tegmark:2006az, Ross:2006me, guzzo, daAngela:2006mf, McDonald:2004xn, Caldwell:2007cw,Dore:2007jh,Giannantonio:2009gi,Mortonson:2009hk,Reyes:2010tr,Daniel:2009kr,Daniel:2010ky,Daniel:2010yt,Bean:2010zq,Zhao:2010dz} could be reinterpreted using this formalism. We believe that this method is useful for two reasons. On the one hand, while it is not conceptually different from fitting the parameters of the 
GR+$\Lambda$CDM model, it enables straightforward, separate tests of two of its properties. On the other, the Figure of Merit we have introduced allows one to compare different surveys on an equal footing, and to easily predict the combined power of different surveys in detecting deviations from the concordance model, which is a new feature of this approach.

The method can also be extended to different observables. For example, in non-standard cosmological models, the scale dependence of the growth factor $D(k,a)$ and its logarithmic derivative $f$ can be completely different. As a result, a simultaneous measurement of these two quantities would be another model-independent test of the concordance model. Galaxy redshift surveys can in principle measure both, since the redshift distortion is sensitive to $f$, but the power spectrum itself is sensitive to the growth factor $D$;
the latter can also be measured via weak gravitational lensing \cite{Massey:2007gh,Fu:2007qq,Zhao:2008bn,Zhao:2009fn}, and galaxy clusters counts \cite{Rapetti:2009ri}. 
We defer this extension to another paper \cite{AG_inprep}. \\

We warmly thank Nicholas Bond, Amir Hajian, Gary Hill, Karl Gebhardt, Eiichiro Komatsu, and David Spergel for useful comments, and Donghui Jeong and Masatoshi Shoji for enlightening discussions. We also thank the referee, Eric Linder, for many suggestions which helped us improve the manuscript.

\end{document}